\documentclass[pdflatex,sn-mathphys]{sn-jnl}

\jyear{2022}%

\theoremstyle{thmstyleone}%
\theoremstyle{thmstyletwo}%

\theoremstyle{thmstylethree}%

\usepackage[utf8]{inputenc}

\begin{document}

\title[Interaction of plasma and magnetic fields in the lower solar atmosphere]{Interaction of convective plasma and small-scale magnetic fields in the lower solar atmosphere}

\author[1]{\fnm{Santiago} \sur{Vargas Domínguez}}\email{svargasd@unal.edu.co}

\author[2,3]{\fnm{Dominik} \sur{Utz}}\email{Dominik.Utz@Kepleruniklinikum.at}


\affil[1]{\orgdiv{Observatorio Astronómico Nacional}, \orgname{Universidad Nacional de. Colombia}, \orgaddress{\city{Bogota}, \country{Colombia}}}

\affil[2]{\orgdiv{Institute of Physics, Faculty of Science}, \orgname{University of South Bohemia}, \orgaddress{\city{\v{C}eské Bud\v{e}jovice}, \country{Czech Republic}}}

\affil[3]{\orgdiv{Neuroimaging Science and Support Center - NISSC, Institute of Neuroradiology}, \orgname{Kepler University Hospital}, \orgaddress{\city{Linz}, \country{Austria}}}


\abstract{In the following short review we will outline some of the possible interaction processes of lower solar atmospheric plasma with the embedded small-scale solar magnetic fields. After introducing the topic, important types of small-scale solar magnetic field elements are outlined to then focus on their creation and evolution, and finally end up describing foremost processes these magnetic fields are involved in, such as the reconnection of magnetic field lines and the creation of magneto-hydrodynamic waves. The occurrence and global coverage in the solar atmosphere of such small-scale phenomena surpass on average those of the more explosive and intense events, mainly related to solar active regions, and therefore their key role as building blocks of solar activity even during the weaker phases of the 11-year solar cycle. In particular, understanding the finest ingredients of solar activity from the lower to the upper solar atmosphere could be determinant to fully understand the heating of the solar corona, which stands out as one of the most intriguing problems in astrophysics nowadays.}


\maketitle

\section{Introduction}

The surface of the Sun, known as photosphere, displays a plethora of features at several spatial and temporal scales, mostly related to the existence of convective flows, which in turn transport plasma and magnetic fields from the solar interior. Sunspots are the most prominent and best studied manifestation of the emergence of the most intense magnetic flux, and their appearance/disappearance in the photosphere is directly related to high/low activity of the Sun, being the basis for the discovery and characterization of the solar cycle.

From the visible sunspots or groups of them, to the finest magnetic elements, studying the evolution of magnetic fields embedded in the photospheric plasma is essential to understand multiple other processes occurring from the lower to the upper solar atmospheric layers, which are responsible for phenomena involving energy release and ultimately have implication on space weather.

Concerning the solar plasma, the photospheric one can be described theoretically well by a single fluid magnetohydrodynamic (MHD) approach due to the comparatively large density and strong collisional coupling between ions and neutrals. This allows for full fledged computer simulations due to the still manageable computational power needed for such simulations \citep[e.g.,][]{2005A&A...429..335V}. Contrary to the photosphere, the corona can be described by a fully ionized plasma changing the theoretical description dramatically \citep[e.g., usage of kinetic approaches during flare events to describe particle acceleration processes][]{2011ApJ...737...24B,2021PhPl...28i2113Z}. The layer in between -- chromosphere --  is the most complicated one to theoretically describe in full detail, as the transition from a weakly ionized plasma to a fully ionized one is occurring in this region. Due to the lower density in the chromosphere, compared to the photosphere, the non-ideal (multi-fluid) effects are more pronounced.  Thus the approach of multi-fluid MHD is heavily utilized nowadays for this layer \citep[][]{2012ApJ...753..161M}, although the pure ideal MHD theory has been very successful in explaining lots of observational phenomena in the solar corona and especially the photosphere.

From the observational point of view, the photosphere is characterized by the presence of flow convective patterns at multiple spatial and temporal scales, being the so-called granulation, mesogranulation, and supergranulation, the more studied ones, from lower to greater scales. Granular cells are characterized by diameters of 100 km, vertical flows of 1 km~s$^{-1}$ and liftimes of 0.2 hr. Mesogranular cells have diameters of 5000 km, vertical flows of 60 m~s$^{-1}$ and lifetimes of  3 hr. Supergranular cells exhibits diamaters of 32,000 km, horizontal flows of 400 m~s$^{-1}$ and lifetimes of 20 hr. While granulation and supergranulation are definitely present, well known and confirmed numerous times, the origin and existence of mesogranulation, also known as family of granules, is still under debate, as summarized in \citep[][and references therein]{2003ApJ...597.1200R}.

Correlation tracking techniques have been extensively used since the 80s to study the evolution of such patterns \citep[][and references therein]{1988ApJ...333..427N,1989ApJ...336..475T,2012A&A...537A..21P}. In particular, the evolution of granules, with their typical expansion and fragmentation, and interaction with embedded small-scale magnetic fields, have been studied in order to characterize the disturbance of the granulation pattern and the accumulation of magnetic fields in intergranular regions
\cite{2003A&A...407..741D, 2011ApJ...735...74I}. The grouping of magnetic fields at different sizes, from small-scales, such as Magnetic Bright Points \citep[MBPs;][]{1973SoPh...33..281D,2004A&A...428..613B}, to large-scale magnetic concentrations like sunspots or Active Regions  can be detected in the photosphere, and have been matter of observational and theoretical studies. While sunspots have been widely observed for centuries, small-scale magnetic features were elusive until the advent of high-resolution solar instrumentation. Such compact and minuscule magnetic fluxes are thought to constitute building blocks of solar activity \citep[see, e.g.,][]{1987ARA&A..25...83Z}, considering that they are, for instance, more ubiquitously distributed in the Sun \citep[see, e.g.,][]{2013SoPh..284..363U} and the fact that exist even during the weakest phases of the solar cycle \citep[][]{2016A&A...585A..39U}, therefore playing an important role in the overall layout of the star \citep[for detailed analysis of the solar cyclic behaviour as well as latitudinal distribution of small scale fields as observed by the Solar and Heliospheric Observatory (SoHO) with the Michelson Doppler Imager (MDI), see][]{2012ApJ...745...39J}.

\section{Small-scale solar magnetic fields}
The solar surface is spreckled with small-scale solar magnetic field elements which are more or less organised. We can differentiate between the regions of occurrence in network magnetic field elements and intranetwork fields. Network regions comprise generally the larger and stronger small-scale elements being characterized by their vertical magnetic field often reaching kG. Intranetwork fields are found inside the magnetic network roughly corresponding to the boundary of the supergranular cells and being, contrary to the network fields, frequently orientated horizontally and weaker, with field strengths of about 100 G. In the following sections, some of the important types of small-scale solar magnetic field elements are introduced.

\subsection{Pores}

\begin{figure}[htbp]
\centering
\includegraphics[width=1.\textwidth]{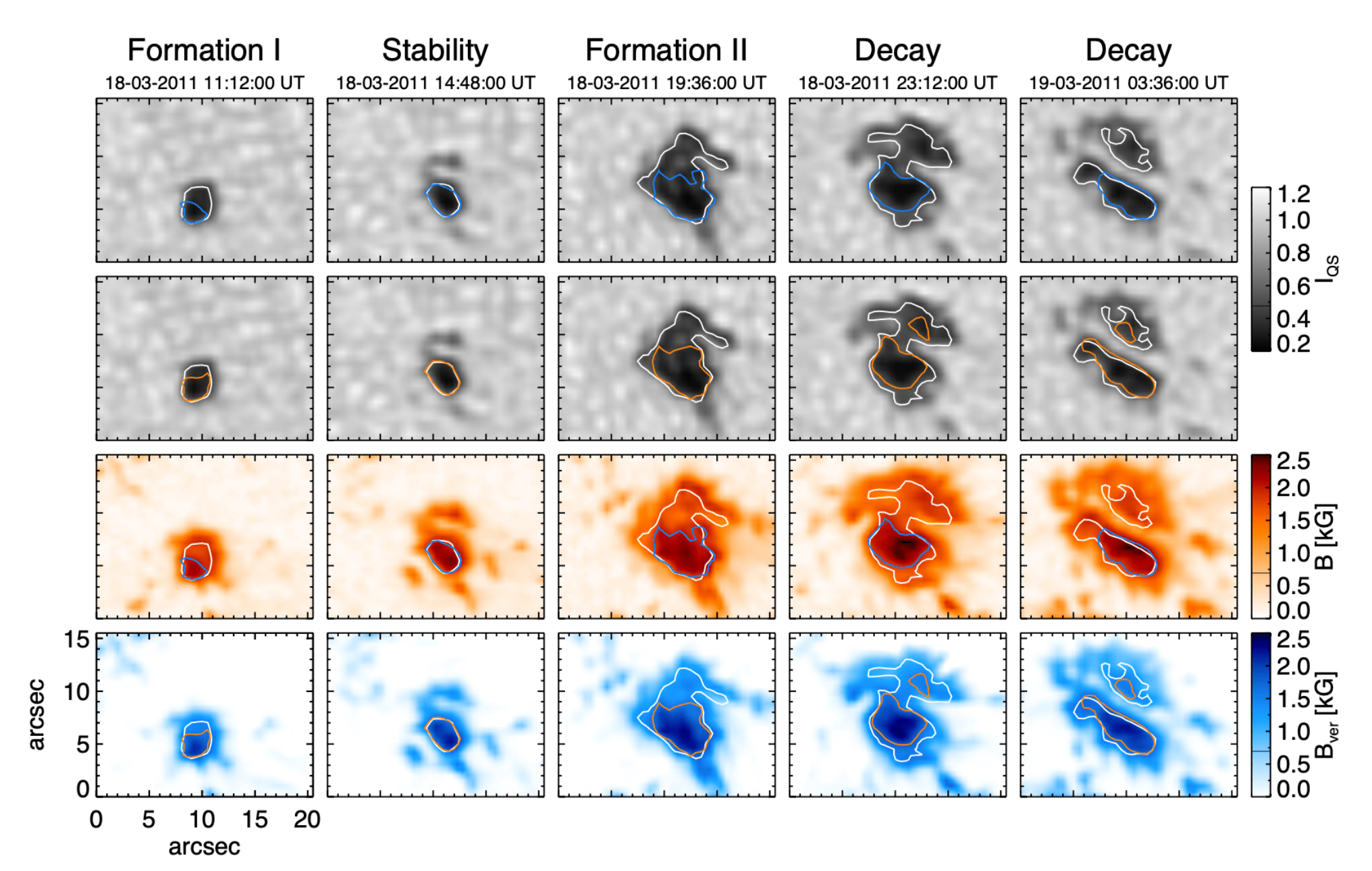}
\caption{Sequence of evolutionary stages of a pore during its lifetime as presented by \cite{2021A&A...649A.129G}. Images show intensity maps with white contours for $I_c = 0.55 I_{QuietSun}$, and magnetic field strength contours in blue for B = 1921 G (top row),  vertical magnetic field contour of  1731 G in orange (second row). Maps of magnetic field strength are shown in the last two rows, using the same contours code as in first and second rows, respectively.} 
\label{fig:fig1}
\end{figure}

Pores are visible structures in the photosphere, characterized by purely dark areas, as “naked” sunspots without a penumbra \citep{2021A&A...649A.129G}, as displayed in Fig.~\ref{fig:fig1}. The generation of pores is a consequence of the inhibition of convection once magnetic field emerges from the boundary between radiative and convective zones (tachocline) due to buoyancy driven instabilities, and crosses the photosphere, thus reducing the transfer of heat from lower layers, making them look dark. Pores exhibit diameters in the range from 1000 to 6000 km \citep[for detailed statistics, see, e.g.,][]{2015ApJ...811...49C}, and can harbour smaller structures produced by the convection not being fully suppressed by the emerging magnetic field, such as umbral dots \citep[e.g.,][]{2006ApJ...641L..73S} and light bridges \citep[][]{2013A&A...560A..84S} commonly known and found in larger sunspots. For many years solar pores were the finest coherent magnetic structures that were able to be studied rigorously in the photosphere while even smaller magnetic field concentrations have been, to a major fraction, characterised only phenomenologically via photometric observations (e.g., bright filigree observations). However, solar pores are, for the current high-resolution solar physicists, just the starting point for going into much smaller spatial scales to discover new exciting phenomena.

Pores seem like a perfect laboratory to study the interaction and evolution of almost vertical emerging magnetic fields with the convective pattern around, considering that, different from sunspots, pores do not exhibit a formal penumbra \citep[see,][]{1964suns.book.....B,2007A&A...474..261C}, which is generated by more inclined magnetic fields. Pores can be therefore regarded as the first step prior to the evolution of a sunspot although not all pores will eventually evolve into sunspots but rather disintegrate and vanish \cite{1992Natur.356..322M, 1997A&A...328..682S}. The relatively simple configuration of the magnetic field in pores represents an appropriate scenario to probe the interaction between the magnetic field and the moving plasma.

There are multiple studies dealing with the fine structures in and around pores, which reveal a complex interplay between the emerging magnetic flux and the convective plasma motions \cite{1999ApJ...511..436S,2010A&A...516A..91V,2017A&A...600A.102E,2022A&A...660A..55X}. In some cases, a transitory penumbra can be formed extending outwards from some regions of the pore. A number of bright dots can also be evidenced inside the pore, resembling  the umbral dots (UD) detected in sunspots' umbra but being more dynamic, brighter and with longer lifetimes, mainly due to the pore's weaker magnetic field that inhibits convection less efficiently \cite{1993ApJ...415..832S}. The immediate photosphere around pores is characterized by flows towards them, which can make small granules penetrate the pore's border similar to UD, with average lifetimes of $\sim$6 min, as studied in \cite{1992SoPh..140...41W} and more recently in chapter 6 of \cite{2009PhDT........78V}. In general, the elongated small-scale features are signatures of more inclined magnetic fields at the border of the pore \cite{2021A&A...649A.129G}.

\subsection{Single flux magnetic fibrils}

\begin{figure}[htbp]
\centering
\includegraphics[width=1.\textwidth]{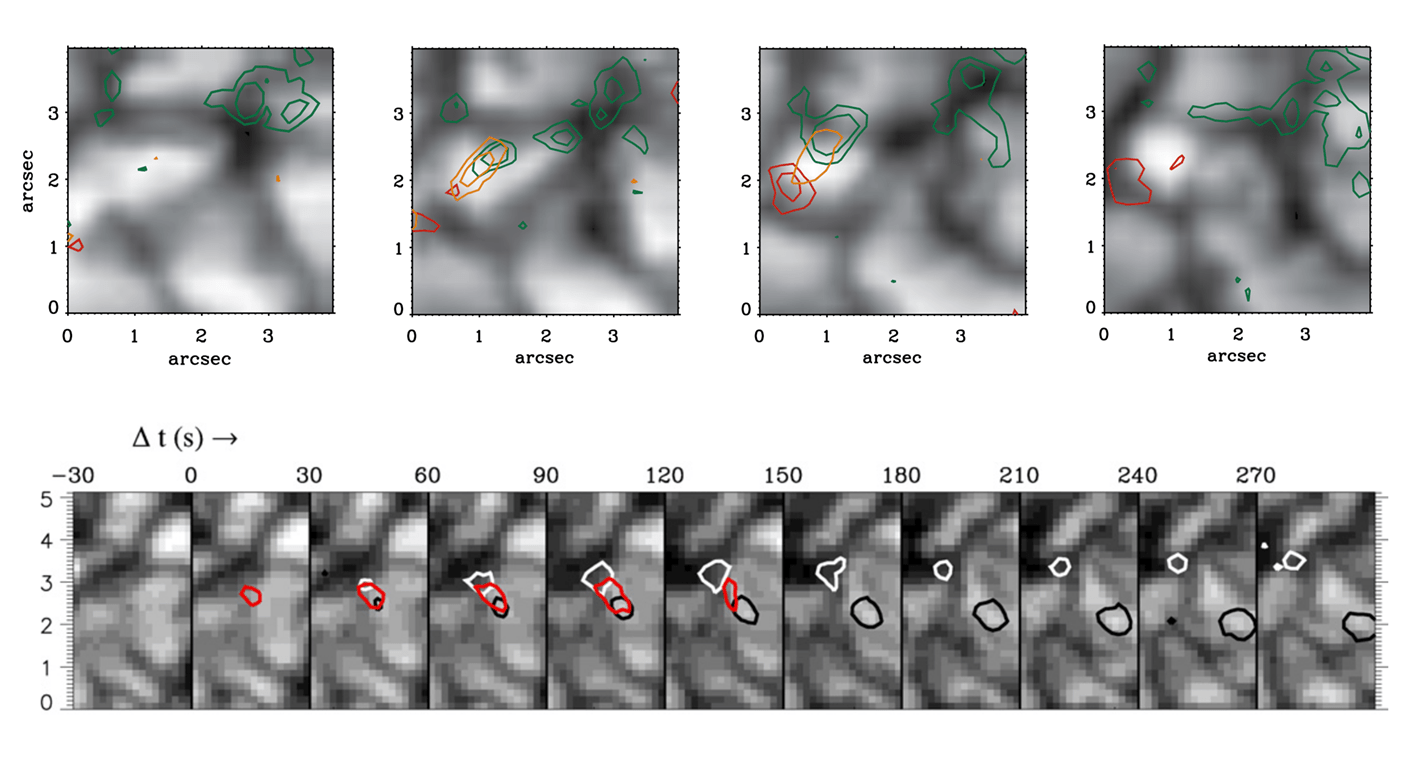}
\caption{Observational evidence of small-scale magnetic loops emerging in the solar photosphere. Top row: Images taken from \cite{2007ApJ...666L.137C} with frames separated by 125 s displaying the solar granulation pattern with overlying contours in red, green, orange for positive circular, negative circular, and linear net polarization signals, respectively. The frames evidence the spatial correlation between the emerging magnetic flux and the location of the granule. Bottom row: Images of the solar quiet granulation pattern taken from \cite{2009ApJ...700.1391M}, with red contours representing linear polarization signals, and black and white contours for circular polarization ones. Axes are in arcsec.} 
\label{fig:fig2}
\end{figure}

Detecting and understanding the behaviour of small-scale magnetic elements needs the combination of both, detailed observations and advanced numerical simulations. For nearly three decades, since the first pioneering observational work of \cite{1996ApJ...460.1019L} and several following works \cite{2007ApJ...666L.137C,2010A&A...511A..14G, 2009ApJ...700.1391M} using spectropolarimetric data at very small spatial scales of less than 2 arcsec (Fig.~\ref{fig:fig2}), it is now well-established the presence and overall action of small-scale individual magnetic loops emerging through the quiet solar atmosphere. The magnetic $\Omega$-loops, which emerge from beneath the photosphere, leave an imprint in the  spectropolarimetric maps. While rising up, the more horizontal segment of the loops (apex) is the first part detected by linear polarization signals, subsequently followed by the more vertical field, which is evidenced by a pair of opposite-polarity circular polarization signals, as displayed in the sequence of frames in the lower panel in Fig.~\ref{fig:fig2}). Recent observations \cite{2017ApJS..229....3C} and numerical simulations \cite{2018ApJ...859L..26M} have evidenced the appearance of magnetic field patches, composed of bunches of fibrils, which cover single granules, exhibiting mainly aligned and horizontal magnetic fields, with the corresponding footpoints rooted close to the granular border. Some of them are low-lying loops which can rapidly submerge and disappear without any detectable action above a few hundred kilometers, but others rising further up can produce brightnings in the lower and upper chromosphere, meaning they are interacting with plasma and background fields at those heights, as will be commented in Section~\ref{sec:magrecon}. It is now clear that a significant fraction of the quiet Sun is populated by magnetic fields. By increasing the integration time of the spectro-polarimeter of the Hinode mission \citep{2007SoPh..243....3K}, to larger and larger values, it was shown that the fraction of the photosphere occupied by magnetic fields increases until the point that one can detect magnetic fields practically over the whole field of view \citep[][]{Bellot_Rubio_2012}. Weak intranetwork fields were found to be mostly horizontally orientated \citep[e.g.,][]{1996ApJ...460.1019L}, and therefore, our understanding of the quiet Sun changed dramatically over the course of the last decades. From a magnetic atmosphere organized in active regions, a magnetic network, and magnetic void regions in between, the picture evolve into a more complex magnetic atmosphere full of small-scale highly dynamic field elements.  This new way of conceiving the solar atmosphere highlights the importance of the small-scale solar magnetic fields for implications on the overall magnetic configuration of the Sun as well as for further implications on the overall solar activity.

\subsection{Magnetic Bright Points}
\begin{figure}[htbp]
\centering
\includegraphics[width=1.\textwidth]{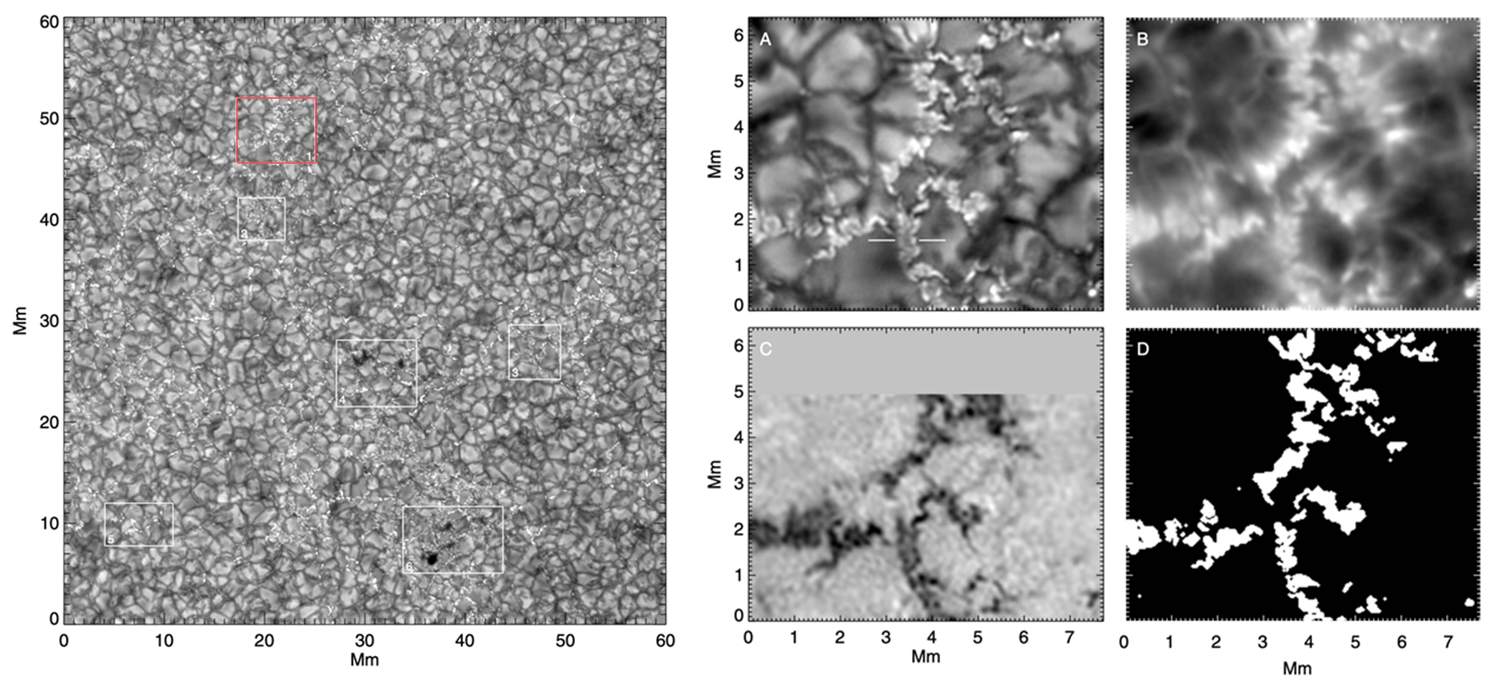}
\caption{Analysis of small-scale solar magnetic elements presented in \cite{2004A&A...428..613B}, showing the AR 10365 observed in G-band (430.5 nm) on 25 May 2003. The boxes with numbers highlight regions of interest inspected by these authors. In particular the region framed in red is further presented in the right panels, displaying G-band filtergram (A) Ca{\footnotesize II}~H (396.8 nm) filtergram (B), Fe{\footnotesize I} (630.25 nm) magnetogram with  with a linear scaling of flux density from $-$1177 to $+$478 Mx.cm$^{-2}$ (C) and a binary mask of the G-band emission (D).}
\label{fig:fig3}
\end{figure}

Magnetic Bright Points (MBPs) have been investigated since the 1970s
\citep{1973SoPh...33..281D} when they were discovered in the intergranular lanes between the photospheric granules. Later their shapes were described in more detail as some MBPs form chain or ribbon-like structures, a kind of magnetic sheet,  while others are situated roundish around granules looking like a flower pattern \citep[see,][]{2004A&A...428..613B}, as shown in Fig.~\ref{fig:fig3}.

In the recent years, it has become more and more common to study the magnetic nature and topology by means of spectro-polarimetric investigations and inversions \citep[][]{2010ApJ...723..787V,2014A&A...568A..13R,2019A&A...630A.139K,2020A&A...633A..60K,2020ApJ...900..130H}. 
MBPs are features in the lower solar atmosphere with an increased brightness compared to their surrounding. Their relative brightness increase is due to the hot wall effect in combination with the possibility to increase the optical depth into the solar atmosphere. After the formation of MBPs, via the most commonly assumed convective collapse process (to be discussed in full detail in a following chapter), MBPs exist as very strong and partly evacuated magnetic field concentrations. To keep the structure evacuated the horizontal inward pressure of the surrounding plasma is balanced by the magnetic pressure of the contained strong magnetic field itself. However, strong vertical upflows might happen in MBPs leading to their destruction \cite{2001ApJ...560.1010B}. The mentioned evacuation of MBPs leads to a lower line formation height compared to its surrounding which now comprises deeper and hotter parts of the photosphere causing an increase in intensity. In addition, the hot walls of the magnetic ``bore hole'' can contribute \cite[especially when observations are taken closer to the solar limb,][]{2004ApJ...610L.137C,2004ApJ...607L..59K,2014A&A...566A..11T}. 

MBPs are mostly studied in the photosphere with the help of the G-band filter. The G-band is a spectral region containing a vast number of CH molecular lines. Thus, using a filter centered at 430.5 nm containing these lines is advantageous as the contrast of small scale magnetic features is increased. This is due to a heat-dissociation effect \cite[e.g.,][]{2001ApJ...555..978S}. As the light emitted within a magnetic structure originates from deeper atmospheric layers due to a local evacuation caused by the magnetic pressure, more of the absorbing CH molecules are dissociated and hence the absorption becomes weaker increasing the intensity of the structure further when compared to normal continuum \citep[all the details can be found in][]{2001A&A...372L..13S,2003ApJ...597L.173S}. Besides of the G-band, also shorter wavelength filters are partly employed such as the CN band head filter at 388.35 nm, showing the features with an even higher contrast and slightly improved resolution under perfect conditions (low straylight contribution, good seeing conditions) \citep{2005A&A...437L..43Z,2014A&A...568A..13R}. In addition to the photosphere, MBPs can also be identified in the chromosphere \cite[for a recent work, see, e.g.,][]{2017ApJ...851...42X}, and a special focus is given here on the connection between the two layers and wave phenomena \cite[][]{2013A&A...549A.116J,2017ApJ...840...19S}.

MBPs can be seen as the visible bright cross-section of strong small-scale vertical magnetic flux tubes. Their magnetic field strength is generally thought to be in the range of kG, often reaching values of about 1.3 kG \citep[see][]{2007A&A...472..607B,2019MNRAS.488L..53K}, although, in the observations  the inverted magnetic field depends a lot on the inversion routine. These are generally not too stable and lead to significant ambiguity in the magnetic field measurements. Nevertheless, \cite{2003ApJ...597L.173S} suggest that the Stokes Inversion based on Response functions (SIR) inversions show the magnetic field up to 1.9 kG. Furthermore, in the simulations, which are not suffering from inversion ambiguities and resolution effects, the magnetic field in MBPs can be even higher than that \cite[e.g.,][]{2014A&A...568A..13R}.

The size of MBPs are observationally found in the range of a few hundred km \citep[$<$300~km;][]{1983SoPh...87..243M,2004A&A...422L..63W,2009A&A...498..289U,2014RAA....14..741Y,2021ApJ...911...32X} down to the resolution limit of the most modern telescopes of 40 to 70 km \cite[][]{2010ApJ...725L.101A}. The actual shape of the size distribution of MBPs is still under debate \citep[e.g.][]{2009A&A...498..289U,2010ApJ...725L.101A} with an ever broader range of evidence pointing to log-normal distributed size distributions \citep[e.g.,][]{2010ApJ...722L.188C,2022A&A...657A..79B}. 

MBPs have generally a very short lifetime in the range of a few minutes \citep[e.g.,][]{2010A&A...511A..39U,2018ApJ...856...17L} as indicated by recent research. Older papers give values also in the range of 10 to 20 minutes \citep[e.g.,][]{1983SoPh...85..113M,2008ApJ...684.1469D}. This can be due to manifold reasons of which the most likely one are different instrumentation and methods to extract the features on the one hand and different selected features on the other hand, e.g., by manual pre-selection of the data sets (active/quiet Sun) and/or, especially in older studies, by a biased visual inspection and selection of the scientist. In former studies, and with lower cadence data, authors most likely biased their analysis to longer features as they were easier to observe, while on the contrary in modern studies, with high resolution telescopes, authors tend to prefer isolated MBPs belonging most likely on intranetwork fields which just might live shorter compared to MBPs related to stronger network magnetic fields. Another problem arises in the definition of lifetime as MBPs are highly dynamic and can split and merge on a regular basis. This fact brings up to the researcher the problem of finding the best considerations for the definition of lifetime for such ever changing features. Finally, the higher cadence in the data means also that the quality of an automated detection algorithm needs to be sufficiently good not to terminate prematurely a time sequence of an evolving MBP by missing one time instance and breaking a sequence into two shorter sequences. Overall, the findings can still be seen as an indication that there might be two populations of MBP features. This speculation is now even more backed up by similar findings in the magnetic field strength distribution \citep[][]{2013A&A...554A..65U,2019MNRAS.488L..53K} as well as in the size distribution \cite{2022A&A...657A..79B}. The distributions of both defining characteristics of MBPs show a bimodal behaviour with two distinct characterising values for the size as well as the magnetic field strength. Such a bimodal parametrization might be related to the occurrence of MBPs which can be found in the network as well as in the intranetwork region possibly giving rise to two populations of MBPs. 

They are very dynamic, often undergoing splitting and merging events, especially when they are part of an enhanced magnetic network. Their horizontal velocities are around 1 km~s$^{-1}$ \citep[e.g.,][]{2003ApJ...587..458N} sometimes reaching velocities as high as 4 to 6 km~s$^{-1}$ \cite{2011ApJ...740L..40K}, which is important for the generation of MHD waves \citep{1993SoPh..143...49C}. Their movement can be overall described by a random walk and they play a role in the diffusion of magnetic fields at the surface of the Sun \cite{1999ApJ...511..932H,2007Ap&SS.312..343S}.

Another interesting topic is the relationship between MBPs and granules. This subject caused more scientific interest some time ago with works like \cite[][]{1989SoPh..119..229M,1992SoPh..141...27M} where the authors studied the location of MBPs and the dynamic caused by the granular flows surrounding them as well as the idea of magnetic compression and granular buffeting. Recent works on this topic nowadays focus less on MBPs as tracers of strong magnetic fields but go more directly into high resolution magnetograms from high cadence magnetographs. An example is the work of \cite[][]{2014ApJ...789....6R} who follow in detail the evolution of a magnetic field patch from the emergence within a granule, the expulsion out of the granule, the agglomeration in a downdraft centre, and finally convective collapse events forming a kG strong MBP signature. Recent research to the interplay of granulation, intergranular lanes, and flux tubes can be found, also in \cite[][]{2019Ap&SS.364..222S}.

While MBPs are tremendously important for small-scale processes and dynamics, as we will see in the next sections, they also play a fundamental role in the structuring of the larger magnetic field topology as can be seen in \cite{2019A&A...629A..22H}. In this work, the authors showed that the footpoints of coronal holes are anchored indeed in small-scale magnetic field concentrations often seen as MBPs.

\section{Generation of small-scale magnetic fields}
To generate magnetic fields one needs an operational magnetic field dynamo. On the global scale this is done in the Sun via an alpha/omega ($\alpha$/$\Omega$) dynamo which can be well simulated nowadays and creates the typical pattern of the sunspot distribution \citep[e.g.,][]{2020LRSP...17....4C}. On smaller scales evidence is mounting for an acting second local dynamo \citep[][]{2015A&A...578A..54T,2017SSRv..210..275B,2018AN....339..127K}. However, magnetic field could be also injected via turbulence from larger scales (sunspots) down to smaller scales during the break up of sunspots at the end of their lifetime. A possible mechanism for this cascading flux injection would be Moving Magnetic Features \citep[MMFs,][]{2019ApJ...876..129L}. MMFs are small-scale magnetic elements which move radially away from decaying sunspots. Finally there are indications that the dynamo process should be even without significant scales as can be seen by a continuous magnetic flux spectrum \citep[][]{2009ApJ...698...75P}.

\begin{figure}[htbp]
\centering
\includegraphics[width=0.9\textwidth]{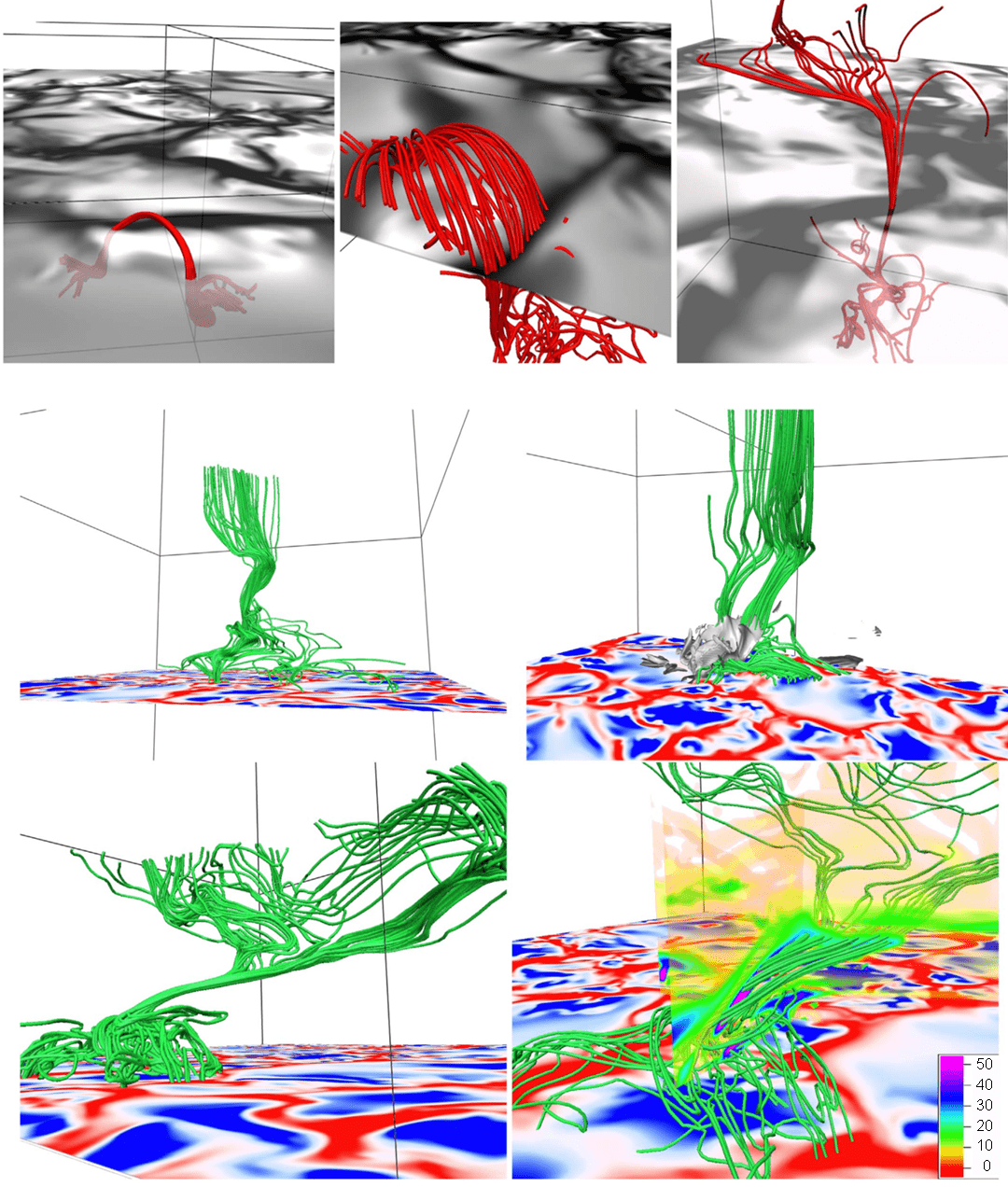}
\caption{Simulations of the solar photosphere and lower chromosphere presented by \cite{2019ApJ...878...40M}. Upper row: Emerging flux tubes (in red) in the photosphere (left panel), flux sheets (middle panel) and flux tube canopies (right). The horizontal map is vertical velocity cut at the photosphere with white/black for 2/-2 km~s$^{-1}$, respectively. Middle and lower rows:  Magnetic field structures in the lower chromosphere, with vertical flux tube concentrations formed at the chromospheric shock fronts (top panels), and flux sheets of up to 64 G (bottom panels). The colorbar shows the field strength in Gauss.}
\label{fig:fig4}
\end{figure}

Recently, there are also first suggestions of a chromospheric small-scale dynamo \citep[][]{2019ApJ...878...40M}. As the authors of the aforementioned work found in their computer simulations that the inflow of magnetic flux from the photosphere into the chromosphere is not sufficient to explain the rapid formation of chromospheric magnetic fields, they suggest that there must be a working mechanism to generate and amplify magnetic fields directly in the chromosphere (see Fig.~\ref{fig:fig4}). In their work they outline also ideas for such mechanisms. To summarize for the photosphere, magnetic flux can be injected from subsurface via flux emergence either by the working global dynamo and possibly on smaller scales by a more shallow secondary small scale dynamo or by shredding from large scale structures.

Flux emergence and the forming of magnetic field elements are well observed for both cases large active regions \citep[e.g, on larger scales][]{2019A&A...622A.168C} as well as single magnetic loops \citep[e.g.,][]{2010A&A...511A..14G}. Sometimes also the emergence happens in the form of magnetic bubbles \citep[]{2014ApJ...781..126O} usually starting within the centre of granules. In a next step, the magnetic field gets expelled from the centre of the granule and starts to accumulate in the inter-granular lanes. Here the magnetic field can agglomerate and amplify up to roughly 400 G. This limit is given due to the build up of magnetic pressure which starts to resist in letting the magnetic field be further compressed by convective motion. From an energetic point of view, the energy stored in the static magnetic field becomes larger than the energy contained in the kinetic movement and thus the corresponding magnetic pressure withstands further compression by the gas pressure \citep[see also][]{2006RPPh...69..563S}. Investigations have revealed that the magnetic fields can travel within the intergranular lanes and are slowly, but steadily, advected into downdraft centres where further magnetic field amplification can happen \cite{2014ApJ...789....6R}. On larger scales the flux transport is dominated by the supergranular flows, and ultimately by meridional flows bringing the magnetic flux to the poles of the Sun and causing there the polar reversals, as part of the global solar magnetic field cycle \citep[see also][]{2014SSRv..186..491J}.

\section{Magnetic field amplification and the formation of the magnetic network}
On the individual magnetic field element level, one can observe magnetic field amplification, which occurs once sufficient field has agglomerated at a downdraft center and an instability, called ``convective collapse'', sets in amplifying the several 100 G strong magnetic field further up into the region of 1 kG and beyond. On larger scales the magnetic network is formed via magnetic flux transport to the boundaries of supergranular cells where the weak intranetwork fields merge to form stronger network magnetic elements.

\subsection{Convective collapse}
\begin{figure}[htbp]
\centering
\includegraphics[width=1.\textwidth]{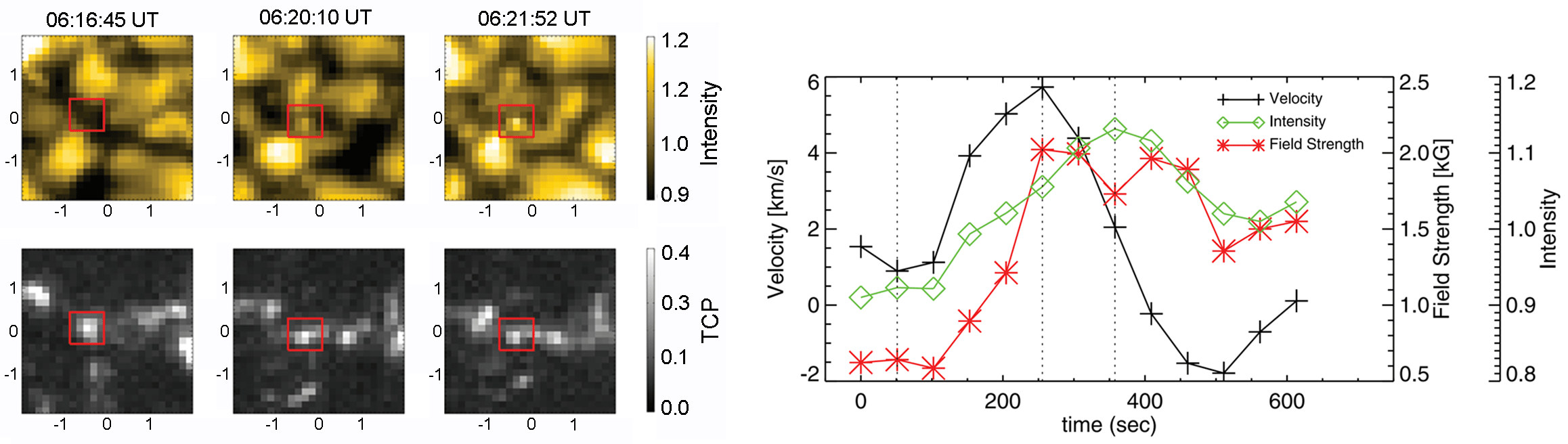}
\caption{Formation of a kG field strength magnetic flux tube as presented in \cite{2008ApJ...677L.145N}. Panels on the left (in arcsecs) display  a sequence (from left to right) before, in the middle, and after the formation of a kG field strength flux tube, respectively.  The red box encloses an intergranular lane where the event is detected. The top row (yellowish frames) show continuum maps derived from raster scans in Fe I absorption line 6302.5 $\AA$, while the corresponding frames below (grayish  frames) show the total circular polarization (labeled TCP in the color bar) of the same absorption line. All spatial units are in arcsec. The plot on the right shows the evolution (starting at  06:15:54 UT on 2007 February 6) of the velocity (black), continuum intensity (green), and field strength (red) observed at the center of the evolving flux tube.}
\label{fig:fig5}
\end{figure}

The convective collapse scenario describes the magnetic field amplification in a magnetic flux tube due to a thermodynamic driven instability and evacuation of the flux tube. The basic scenario states that a sufficient strong magnetic field concentration in the form of a flux tube can start to inhibit the convective influx of new energy in the form of heat from the subsurface layers. Thus, as there is a radiation loss of energy and thus heat from the plasma confined in the magnetic field configuration, the plasma starts to cool down and consequently these now denser and heavier plasma starts to flow down and out of the flux tube. Due to the small-size of the magnetic fields under investigation, e.g. MBPs, this process is very rapid leading to the process named convective collapse as the evacuated flux tube now has not sufficient inside pressure to prevent the magnetic field from compression/collapse. The collapse leads to a strong amplification of the magnetic field, e.g. in the case of MBPs from sub kG fields to kG fields. The process was theoretical described by \cite{1978ApJ...221..368P} and \cite{1979SoPh...61..363S} and later impressively shown by observations from the Hinode mission in a case study \cite{2008ApJ...677L.145N}, (see Fig.~\ref{fig:fig5}). Here the observables (line-of-sight velocity, brightness, and magnetic field strength) followed the theoreticized temporal evolution, namely that a strong downflow is followed by an evacuation of the flux (increase in brightness) and collapse of the magnetic field (strengthening of the magnetic field strength). This result was verified and set on more robust statistical foundations by \cite{2009A&A...504..583F}. Early first indications for the convective collapse process causing the magnetic field amplification are dating back to at least the 90s \citep[]{1996A&A...310L..33S}. Besides, researchers were searching for the effect of the convective collapse in simulations and comparing the magnetic field intensification as observed in the simulations with real observations. This means that they were comparing the typical time-scales (a few minutes), downflow velocities (4 to 10 km~s$^{-1}$) and magnetic field amplification (from a few hundred G to above a kG) and found good agreement between observations and simulations \citep[e.g.,][]{2010A&A...509A..76D}.

Nowadays, the convective collapse scenario is considered the process being responsible for the creation of e,g., MBPs. While not only indicating the convective collapse as a fundamental process for the creation of MBPs \cite{2001ApJ...560.1010B} also gave an idea about the weakening and dispersion of MBPs by a back-propagating shock wave created by the impact of the evacuated plasma on the lower denser atmospheric plasma. These back-propagating shock waves increase the diameters of the flux tube, decrease thus the field strength and are probably a cause for the destruction and short lifetime of intense MBPs undergoing very strong convective collapse processes. Very recently, evidence is showing up for further processes leading to higher magnetic field strengths and thus amplifications in MBPs regardless of the convective collapse process \cite[see][]{2014ApJ...796...79U,2020A&A...633A..60K}.
In the following subsection we will see how on larger scales, and statistically, magnetic flux merging can lead to strong magnetic elements forming the magnetic network pattern.

\subsection{Network magnetic flux}
Network magnetic flux can be lost due to submergence or cancellation with opposite polarity elements (as we will see in the last section) and thus needs to be to be continuously replenished. The research points to the idea that practically the whole amount of magnetic flux in the network needs to be recreated within 24 hours. But, where is all the magnetic flux coming from?
Several different hypothesis were stated, among them that the flux is coming from ephemeral regions \citep[see, e.g, the introduction of][and references therein]{2001ApJ...555..448H}. Recent studies with Hinode, however, show that the small-scale flux emergence within the intranetwork regions seem to be sufficient to create enough magnetic net flux to stabilize the network \citep[see,][]{2014ApJ...797...49G,2016ApJ...820...35G}. The authors of the aforementioned papers investigated an exceptional long time series of the Hinode$/$SOT instrument and tracked the evolution of appearing small-scale intranetwork fields. These magnetic elements drift slowly from the inside of the supergranular structure to its boundary and join there the persistent magnetic network. The estimated rates of newly added flux are comparable to believed disappearance rates and thus seems to be sufficient to explain the appearance and maintenance of the magnetic network.

\begin{figure}[htbp]
\centering
\includegraphics[width=1.\textwidth]{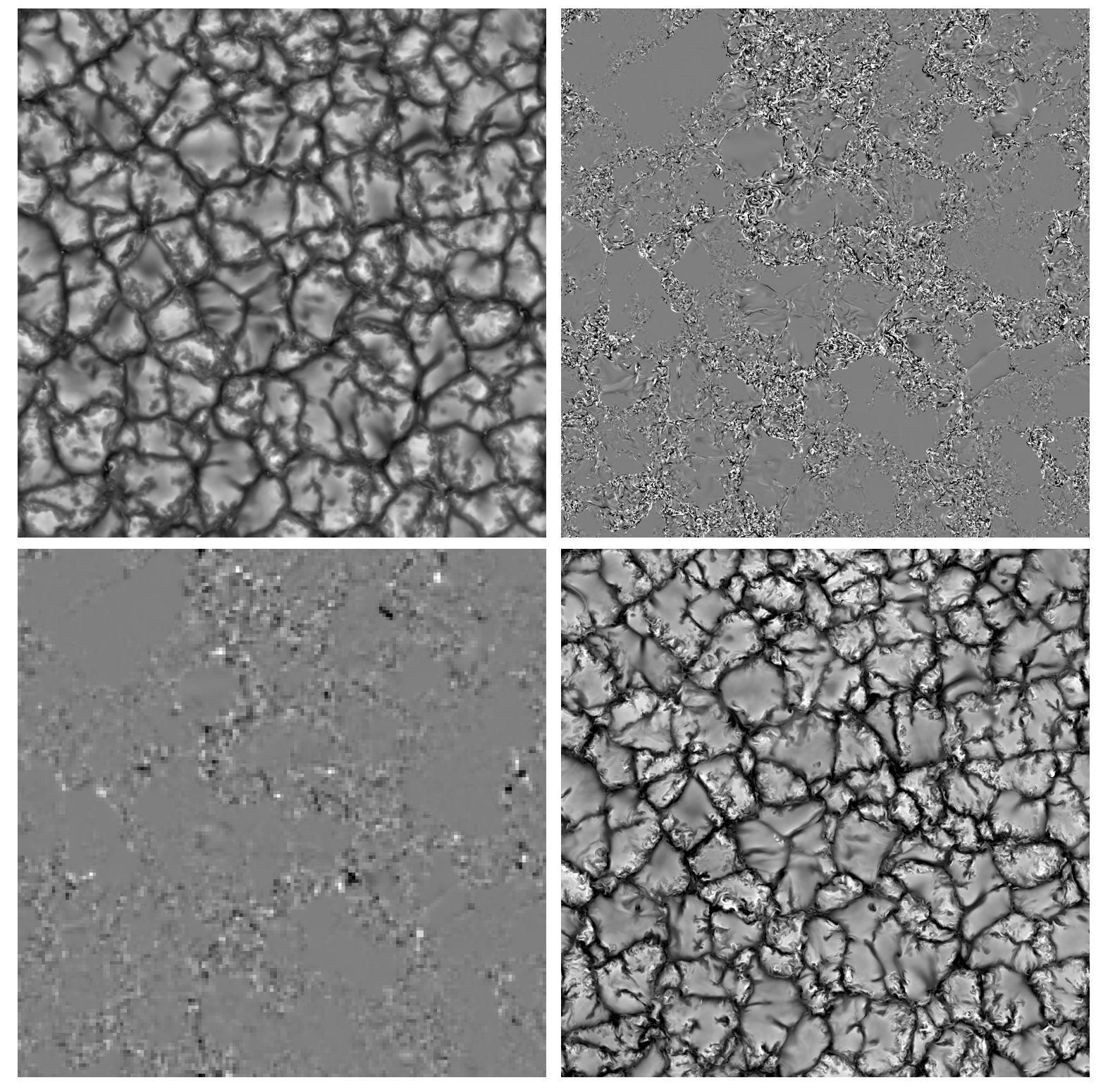}
\caption{Results from a small-scale dynamo simulation with the MURaM code as presented in \cite{2015SSRv..tmp..113B}, showing
from left to right and top to bottom: bolometric intensity,  vertical magnetic field (saturated at $\pm$100G), smoothed vertical magnetic field corresponding to a 50-cm telescope at 630 nm, and the vertical velocity component with brighter/darker
representing upflows/downflows saturated at $\pm$5 km~s$^{-1}$. The computational box has dimensions of 18~Mm $\times$ 18~Mm $\times$ 7~Mm with a grid cell size of 10 km.}
\label{fig:fig6}
\end{figure}

Figure~\ref{fig:fig6}, taken from \cite{2015SSRv..tmp..113B}, shows en example of recent numerical simulations dealing with the analysis of the small scale dynamo. Moreover, the classical salt (positive polarity; white) and pepper (negative polarity; black) spotted image is created. This happens due to the visible prevalence of stronger vertical network magnetic fields, especially in earliest magnetograms with lower resolution.

\section{Plasma - magnetic field interaction}
Multiple processes involving the interaction of plasma and magnetic fields are present in the solar atmosphere. Among the main ones are the buffeting of magnetic fields by the granular motions \citep[e.g.,][]{1992SoPh..141...27M}, the creation of vortex flows \citep[][]{2008ApJ...687L.131B,2018ApJ...869..169G}, and the convective transport of magnetic fields (e.g., on small-scales from the centre of granules to the intergranular lanes and on larger scales from the inside of supergranules to the magnetic network). Except from them, the most important processes are driven by magnetic field reconnection leading to explosive phenomena such as jets, spicules, fibrils, mottles and other outflow phenomena as well as the driving of magneto-hydrodynamic waves. 

\subsection{Magnetic Reconnection}
\label{sec:magrecon}

In plasma physics, it is well known that magnetic field lines are ``frozen-in'' to an infinitely conductive plasma. Magnetic reconnection happens when the frozen-in concept breaks down and magnetic field lines shatter and reorganize, to form a new magnetic topology \citep[see, e.g.,][]{2018A&A...609A.100S}.  In this process huge amounts of energy can be released (i.e., magnetic energy gets converted into heat, and kinetic energy) and particles accelerated \citep[][]{2016ApJ...827...94Z}. In the case of small-scale magnetic fields magnetic reconnection might lead to phenomena like jets \citep[][]{2019ApJ...883..115N} and spicules \cite[]{2009ApJ...702....1H}, as well as explosive events such as Ellerman Bombs \citep[][]{2016ApJ...823..110R}. In the recent paper of \cite[]{2019Sci...366..890S}, the authors report on the observation of small-scale strong magnetic field elements with the Goode Solar Telescope, formerly New Solar Telescope \citep[NST;][]{2010AN....331..620G} at Big Bear Observatory, which is among the leading and highest resolving solar telescopes under operation only outpassed by the recent first light of the Daniel K. Inouye Solar Telescope (DKIST) facility at Mauna Loa Observatory \citep[][]{2021SoPh..296...70R}. 
The authors observed a weak intranetwork magnetic field element approaching a strong opposite network field element. Once the weak element practically touched the stronger network element it started to disappear while simultaneously a spicule was launched to the upper atmosphere. The spicule was subsequently heated to coronal temperatures as observed by the Atmospheric Imaging Assembly instrument onboard the Solar Dynamics Observatory \citep[SDO;][]{2012SoPh..275...17L,2012SoPh..275....3P}.

The authors concluded that they observed magnetic reconnection happening on small-scale magnetic fields in the lower atmosphere which can explain the triggering of spicules and subsequent heating of the quiet Sun upper atmospheric layers.

\subsection{Plasma brightenings, jets and surges}
The magnetic field emerging to the solar photosphere (Emerging flux regions, EFR) can shape the structure of solar atmospheric layers, in particular due to the interaction of the emerging fields and the pre-existing (ambient) fields \cite{1985SoPh...95....3Z,2001ApJ...549..608M,2011SoPh..268..271S}. Magnetic reconnection is the mechanism responsible for most of the small-scale explosive phenomena releasing energy, and mainly visible as brightenings and jets, among others. These phenomena have been proved to be a source of energy and mass injection into higher layers of the solar atmosphere, according to numerical simulations \cite{2005ApJ...635.1299A,2007ApJ...657L..53I,2010ApJ...714.1762P}, and observationally revealed by high-resolution solar telescopes \cite{2013A&A...555A..19G,2012SoPh..278...99V}. The complexity of the real 3D scenario, in which a large population of small-scale magnetic fields with different orientations rise up from the photosphere to the lower chromosphere, and subsequently interact with background fields, can generate a wide variety of jets that contribute to the energy budget and mass of the upper chrosmosphere and beyond.

\begin{figure}[htbp]
\centering
\includegraphics[width=1.\textwidth]{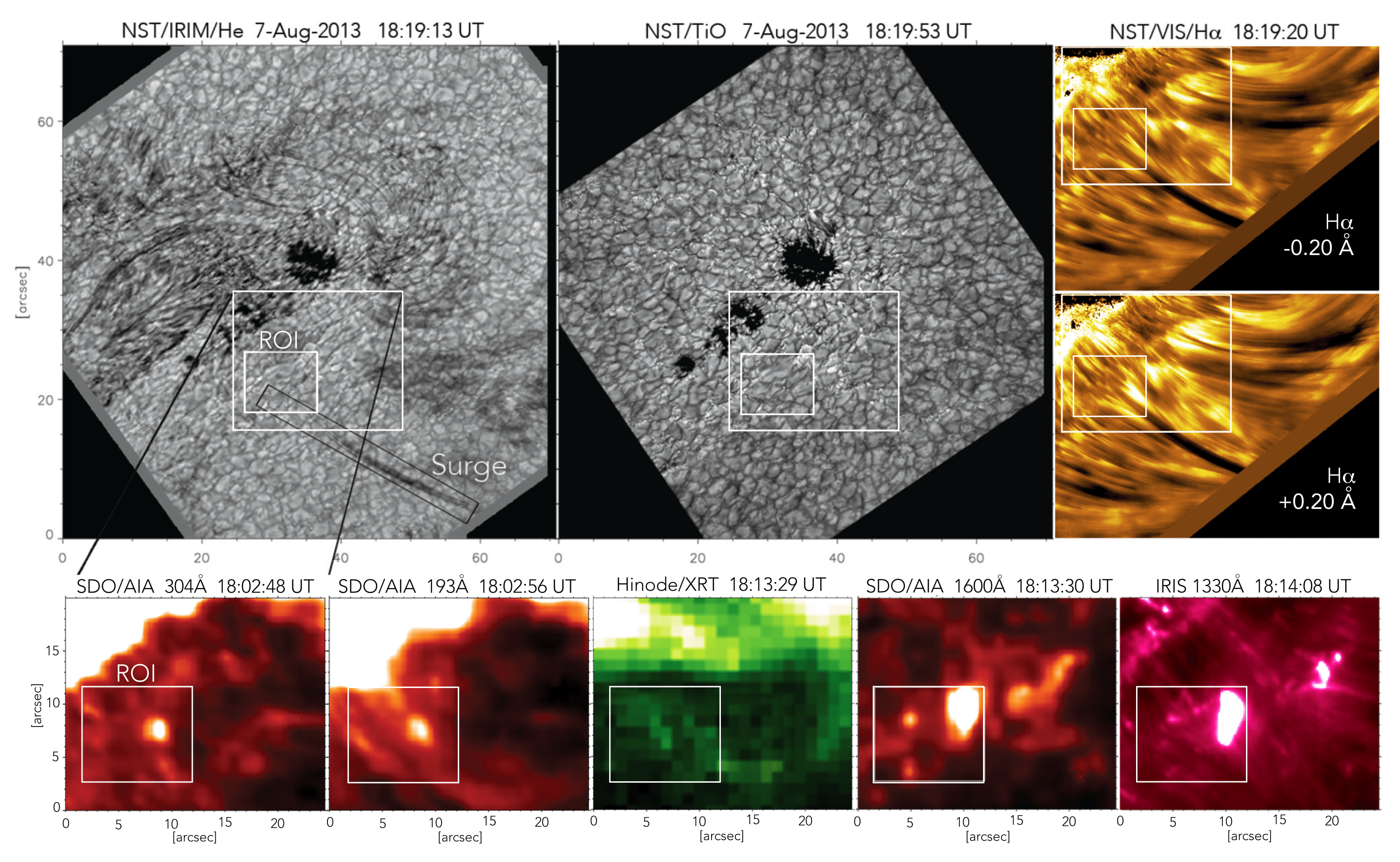}
\caption{Multi-wavelength observations of the solar atmosphere presented in \cite{2014ApJ...794..140V} using data from ground-based (NST) with the instruments Visible Imaging Spectrometer (VIS) and InfraRed Imaging Magnetograph (IRIM), and space telescopes (SDO, IRIS and Hinode).  Panels show images of NOAA 11810 on August 7, 2013.  Upper row: NST/IRIM/HeI image (left panel) displaying a dark-absorbing feature (surge), NST/TiO photospheric image (middle panel), and NST/VIS/H$\alpha$ images in the blue/red wing, as labeled (right panels). Lower row: Chromospheric and coronal images  of the region framed by large white boxes in the upper panels, highlighting the localized brightenings in transition region and coronal heights. All units are in arcsec.}
\label{fig:fig7}
\end{figure}

Figure~\ref{fig:fig7}, adapted from \citep{2014ApJ...794..140V}, shows multi-wavelength high-resolution observations of a small-scale emerging magnetic flux event acquired with the ground-based NST and space-borne solar telescopes; SDO, Interface Region Imaging Spectrograph \citep[IRIS;][]{2014SoPh..289.2733D} and the Hinode mission, detailing the response of different solar atmospheric layers. The interaction of the emerging fields with the overlying field causes high-temperature emission, evidenced as localized brightenings, UV bursts due to plasma heating, as well as the generation of cool surges and hot plasma jets, as shown in the figure.

Explosive small-scale transient episodes, including the so-called Ellerman Bombs (observed in the wings of H$\alpha$) and UV bursts, are now commonly reported through detailed high-resolution observations of the solar chromosphere and transition region \citep[e.g.,][]{2015ApJ...810..145D,2018ApJ...856..127G,2020A&A...633A..58O} and in numerical simulations \citep[e.g.,][]{2008ApJ...687.1373C,2016ApJ...822...18N,2019A&A...626A..33H} dealing with complex 3D scenarios, demonstrating how the reconnection process can provide the energy to heat, and accelerate particles within the plasma.

In a recent observational study \cite{2017ApJ...851L...6R}, the authors furnish compelling evidence for the existence of
plasmoids (magnetic islands), revealed by very dynamic brightenings at sub-arcsec scales, providing support for the interpretation of
intermittent magnetic reconnection driven by the plasmoid instability, previously studied through MHD numerical simulations \cite[e.g.,][]{2013ApJ...777...16Y,2016ApJ...822...18N}.

Further studies should be carried out to establish, and fully understand, the dominant modes of reconnection and the physical conditions responsible for the generation of the wide variety of chromospheric small-scale phenomena, and their implications on the heating of the transition region and corona.

\subsection{Waves}
A hot topic in solar physics is the one related to the generation, propagation, and absorption of waves within the solar atmosphere \citep[e.g.,][]{1999SoPh..186...67E,2003SoPh..217..199T,2005ApJS..156..265C,2013SSRv..175....1M,2015SSRv..190..103J,2022SSRv..218...13N}. The community investigated early on if episodic acoustic waves \cite[][]{1948ZA.....25..161B} could be a possible contributor to the coronal heating problem (under which the maintenance of the million degree Kelvin hot outer atmosphere is summarized) and was able to rule out that possibility as such waves could not penetrate into the higher atmosphere \citep[characterized by the so called cut-off frequency, see, e.g,][]{2016ApJ...819L..23W}. It also appears that pure accoustic waves would not suffice to heat the quiet chromosphere \cite[][]{2007PASJ...59S.663C}. Thus, as the solar atmosphere is permeated by magnetic fields, it is self-evident to look for magnetic waves opening the field of magnetohydrodynamic wave investigations.

One of the important recent topics aims at probing solar pores as magnetic wave guides. The umbral regions of pores typically exhibit 3 mHz oscillations, as consequence of p-modes penetrating the magnetic region, though 5 mHz oscillations have recently been detected in a solar pore \cite{2021A&A...649A.169S}. There are already some observational signatures of resonant MHD oscillations confined within the pore umbra and evidence of propagating MHD-wave activity above pores \cite{2021RSPTA.37900172G}. Wavelet and Fourier techniques, for instance, have been used to detect oscillations in the intensity and the cross-sectional area in pores \cite{2015ApJ...806..132G}.

Considering more compact magnetic fields, due to their vast amount, small-scale magnetic flux tubes, e.g., as seen by MBPs, would represent perfect wave guides to explain the heating of the quiet Sun upper atmospheric layers, if sufficient wave energy can be created, injected, propagated, and finally absorbed in the higher atmosphere \citep[][]{2020A&A...634A..63K}.
Thus the community started already in the 70s and 80s with efforts to model magnetic flux tubes which are nowadays essential as input elements for advanced numerical simulations \citep[e.g,][]{2013MNRAS.435..689G,2018MNRAS.474...77M}.

The problem of the wave propagation then was tackled from the theoretical, observational \citep[see, e.g., the review][]{2013SSRv..175....1M,2012ApJ...744...98C}, as well as computational \citep[][]{2009A&A...508..951V,2011ApJ...727...17F} side, with various studies claiming that small-scale magnetic field elements and the waves which propagate in them, would be sufficient to maintain the quiet Sun´s temperature at higher atmospheric heights. An important part is to understand how the waves ultimately are damped and absorbed, and how the energy can be added to the local atmosphere \citep[e.g.,][]{2021RSPTA.37900172G,2021A&A...648A..77R}. Some recent modelling works study the role of ``non-classical'' ambipolar diffusion on absorption of waves in the small-scale magnetic field concentrations, with variable success \citep[see, e.g.,][and references therein.]{2016ApJ...817...94A, 2016ApJ...819L..11S, 2019ApJ...885...58C}
.

\begin{figure}[htbp]
\centering
\includegraphics[width=1.\textwidth]{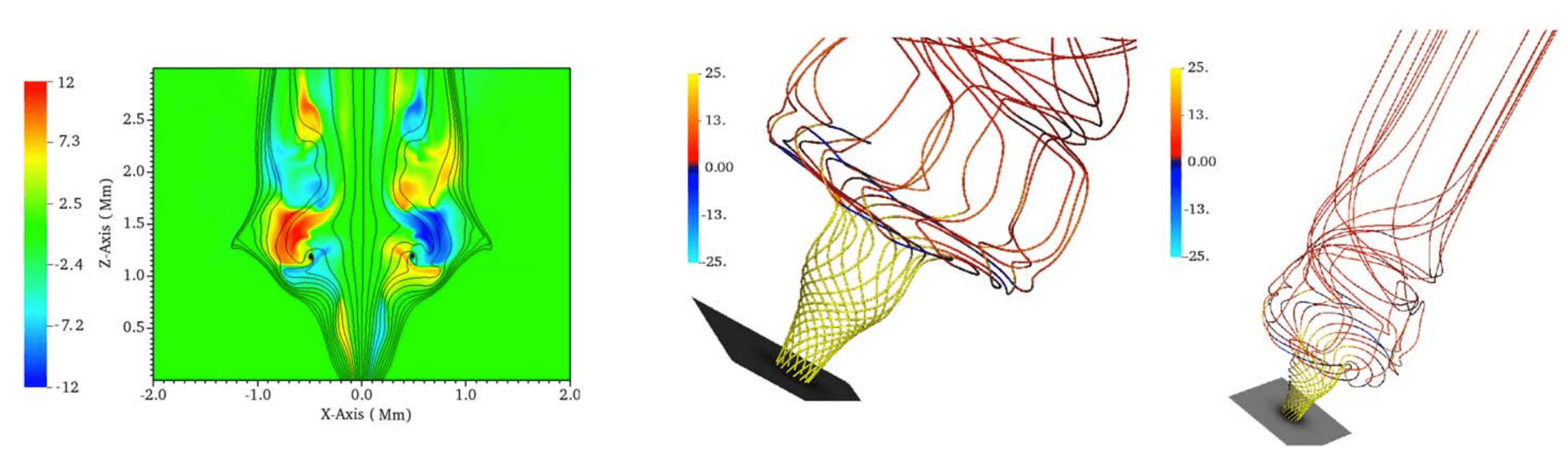}
\caption{Simulation of an flux tube representing an MBP driven by an oscillating vortex flow presented in \cite{2021ApJ...914....8M}. Left panel displays the late-stage dynamics, with the line-of-sight component of the vertical velocity perturbation showing a slice along the y-axis with overplotted magnetic field lines (colorbar in units of km~s$^{-1}$).  Middle and right panels show 3D plots of certain magnetic field lines, with colors according to the corresponding magnitude of the vertical component of the magnetic field, and the bottom boundary displayed in gray panel (colorbars are in units of G). The important part is that some field lines develope negative vertical field strength (blue color) signifying that they are ``switched back'' now. If such field lines can propagate upwards with the disturbance and survive the propagation through the atmosphere they might be a possible source for the magnetic field line switchbacks found within the solar wind stream.}
\label{fig:fig8}
\end{figure}

Besides of this ``classical'' wave studies, the recent discovery of so-called magnetic field line switchbacks within the solar wind by the NASA’s Parker Solar Probe \citep[PSP;][]{2016SSRv..204....7F} mission, further increased the interest also in small-scale magnetic fields as possible source region for such switchbacks. First simulations were not totally conclusive as they have shown that in principle driving of small-scale flux tubes in the lower solar atmosphere can cause the formation of magnetic field line switchbacks, however, they seem not to survive a propagation through the higher solar atmosphere to ultimately join the outstreaming solar wind, as presented in \citep[see][]{2021ApJ...911...75M,2021ApJ...914....8M}. In the latter work, the authors simulate an oscillating vortex extending upwards from the photosphere, as shown in Figure~\ref{fig:fig8}. In the recent years, plasma and magnetic vortices have been of special interest and were tackled  by both observing and simulation approaches \cite[][and references therein]{2008ApJ...687L.131B,2011AnGeo..29..883S,2011MNRAS.416..148V,2021ApJ...915...24S,2022ApJ...928....3A}. The interest in the magnetic vortices arises as they are involved in manifold interesting processes such as the breading of magnetic field lines that could stimulate magnetic reconnection, producing plasma heating, jets, and eruptive phenomena. Besides, they can also be sources of upwards wave propagation. The relationship between vorticity generation, magnetic fields and Alfvén waves has been shown using simulations in \cite{2011A&A...526A...5S, 2012Natur.486..505W, 2013ApJ...776L...4S, 2021A&A...649A.121B}.

Another direction of wave research is aimed to the understanding of torsional Alfvén modes and waves\cite[][]{2002A&A...395..669L,2011A&A...534A..27M,2011AnGeo..29.1029F} as they are possible further agents to heat the higher atmosphere \cite[][]{2021ApJ...909..190S}. Such oscillations were observed at the solar limb in higher atmospheric layers \cite[][]{2012ApJ...752L..12D}. This is due to the fact that the oscillation would cause, perpendicular to the rotation axis of the plasma column, counter-rotating plasma \cite[][]{2012ApJ...759..144T}. Such a plasma would yield a common observable signature of a split plasma column where one side would move to the observer and one side away from the observer. However, for small-scale magnetic fields on the disk such observations would be tricky as the structures are way too small to directly detect the caused motions. Thus, researchers employed up to now mostly indirect methods like measuring line width oscillations caused by the torsional oscillation \cite[][]{2009Sci...323.1582J}.
However, in the recent past now more works are published about the detection of torsional oscillations in the lower solar atmosphere. In the following case the oscillations were detected in observational data within a magnetic pore \cite{2021NatAs...5..691S}, with a kink mode plausible to be an excitation mechanism of the generated Alfvén waves, and are proposed to be a significant contribution to the energy transport in the atmosphere and, moreover, in the acceleration of the solar wind.

\begin{figure}[htbp]
\centering
\includegraphics[width=1.\textwidth]{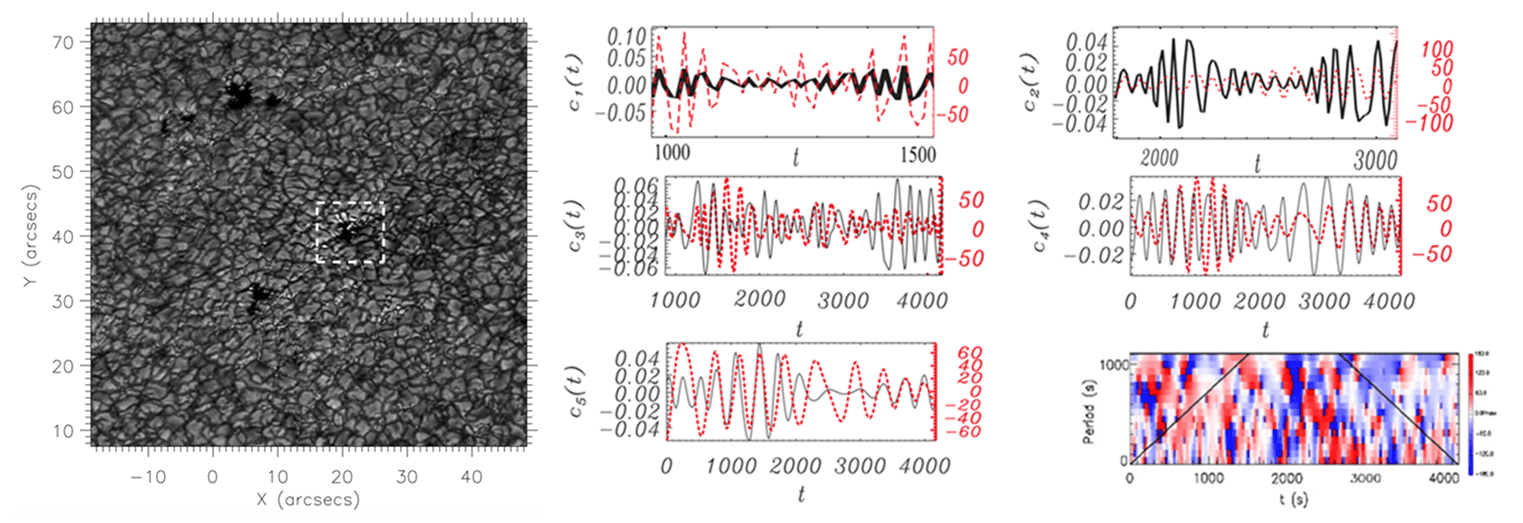}
\caption{Observational analysis of data obtained by the ROSA instrument situated at the Dunn Solar Telescope illustrating the idea of the phase-relationship interpretation for sausage mode waves detected in oscillations on 2008 August 22, as presented by \cite{2011ApJ...729L..18M}. The pore under analysis is boxed in white (left image). Plots on the right show a direct comparison between Intrinsic Mode Functions (IMFs) components of intensity (black solid lines) and pore size (red dashed lines), with he intensity normalized with respect to the background mean value. The size of the pore is shown in pixels). The bottom-right panel displays a wavelet phase plot comparing the pore size and intensity time series.}
\label{fig:fig9}
\end{figure}

From the viewpoint of observations, one can differentiate between studies trying to shed light on waves in magnetic structures by direct evidence of the waves, such as a translation or deformation of the wave guide (like oscillating coronal loops), as well as changes in intensity, and more subtle indirect evidences such as, e.g. a change in the spectral line width. This method was successfully applied to MBPs to show the possibility of torsional Alfvén waves in MBPs \citep[][]{2009Sci...323.1582J}, while kink and sausage magneto-acoustic waves are often detected via displacements, intensity variations or size variations \citep[][]{2013A&A...551A.137M,2013A&A...555A..75M,2017ApJ...840...19S,2017ApJS..229...10J} of the wave guiding magnetic field element. Figure~\ref{fig:fig9}, extracted from \cite{2011ApJ...729L..18M}, shows an observational detection of sausage modes in magnetic pores. Recently, the degree of understanding the observations goes as far as detailed mode identification \cite[even in non-circular shaped flux tubes][]{2021ApJ...912...50A} and differentiation between body and surface type modes \cite[][]{2015A&A...579A..73M,2018ApJ...857...28K}.

Finally, MHD waves are also helpful in revealing physical information about the magnetic field in higher atmospheric layers. Due to the weaker magnetic field the Zeeman effect becomes to difficult to be applied successfully to detect magnetic fields in the higher solar atmosphere. An alternative window is opened by magnetoseismology, the art to understand MHD waves in such detail as to decipher physical quantities (generally, magnetic field strength, density, temperature) of the wave guide \citep[e.g.,][]{2012ApJ...744....5M}. 

From the viewpoint of theory \cite[][]{1983SoPh...88..179E}, the focus is shifting more and more to understand partly ionized wave guides \citep[e.g,.][]{2019AdSpR..63.1472B,2022MNRAS.511.5274A} as this will open new avenues in better understanding and addressing waves in the chromosphere. In addition, there is still development going on to fully understand how waves get damped and energy absorbed \citep[][]{2020ApJ...899..100V}. Properties of sausage waves and their propagation are address from numerical simulations, e.g., initiated at the base of the photosphere in a gravitationally stratified and viscous plasma \cite{2006PhPl...13d2108B}, along a wide flux tube expanding with height \cite{2014SoPh..289.1203P}, among many others.

Finally, we wish to mention that in the recent years the boundary between jets, spicules, fibrils and other mass flow phenomena and waves are starting to dissolve, as there is more and more work contributing to the coupling of the two phenomena. This means, on the one hand, to understand how jets can create waves. Here we wish to mention exemplarily  the work of  \cite{2021ApJ...913...19M}. In this work the authors were reporting on wavelike substructures which were forming within a jet. Thus, it is an example of a jet causing a wave phenomenon.
On the other hand, we mentioned that there are waves causing jet phenomena. In this case we wish to mention as an example the recent work of \cite[][]{2021ApJ...922..118S}, where the creation of jets and outflow phenomena due to waves are discussed.

\section{Removal of magnetic field}
\begin{figure}[htbp]
\centering
\includegraphics[width=1.\textwidth]{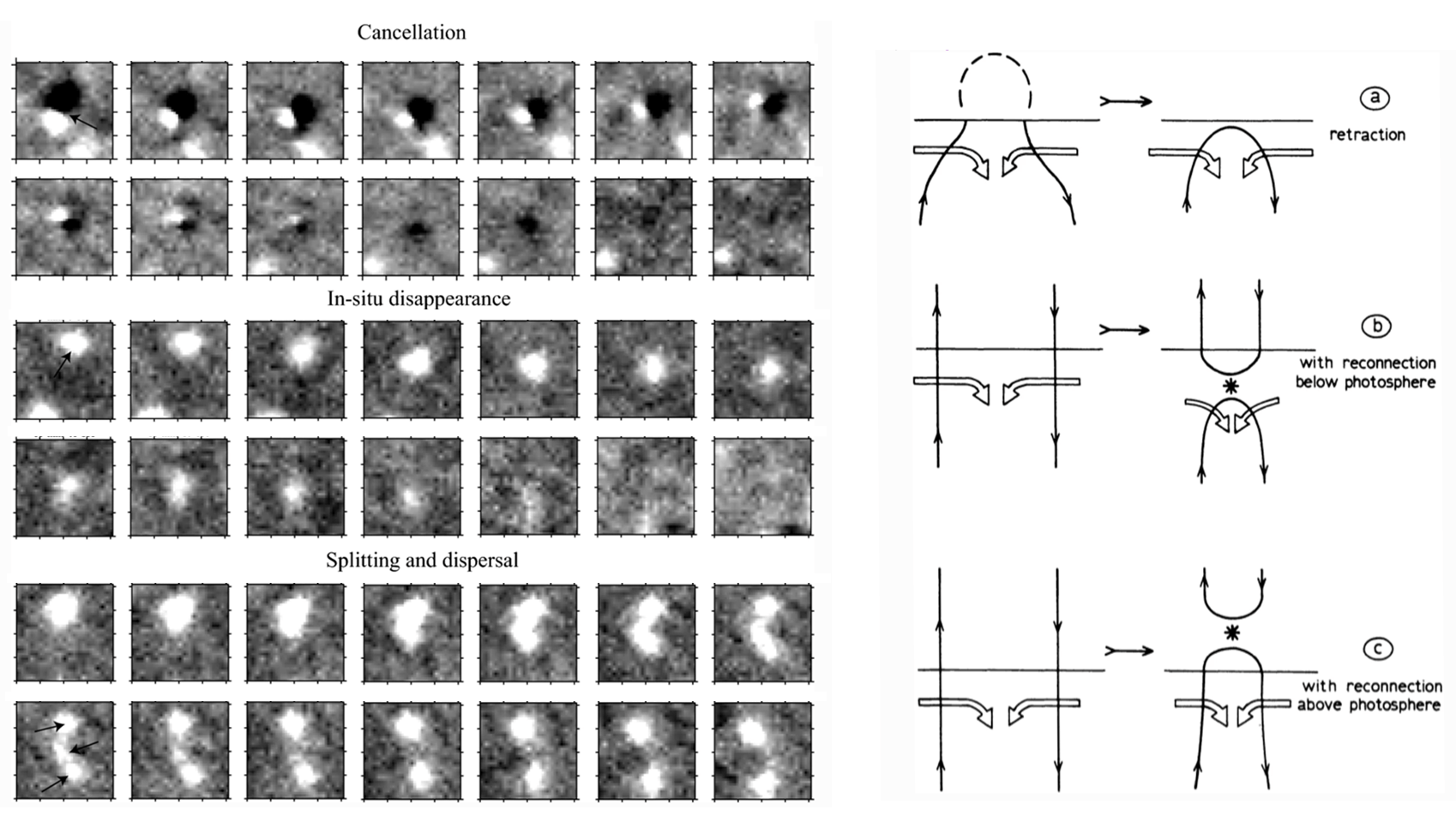}
\caption{Observational evidences for cancellation, disappearance, and fragmentation of magnetic flux features as presented by \cite{2019LRSP...16....1B} through sequence of images (with a cadence of 100 s). White/black represent positive/negative polarity. The magnetograms are saturated to $\pm$~30Mx cm.s$^{-1}$ to highlight the evolution of the magnetic flux. Tickmarks are separated by 1 arcsec. The sketch on the right represents possible cancellation scenarios proposed by \cite{1987ARA&A..25...83Z}, including the submergence (retraction) of existing field lines forming an $\Omega$-loop (a), and prior reconnection of unrelated field lines below (b) of above (c) the photosphere.}
\label{fig:fig10}
\end{figure}

\begin{figure}[htbp]
\centering
\includegraphics[width=1.\textwidth]{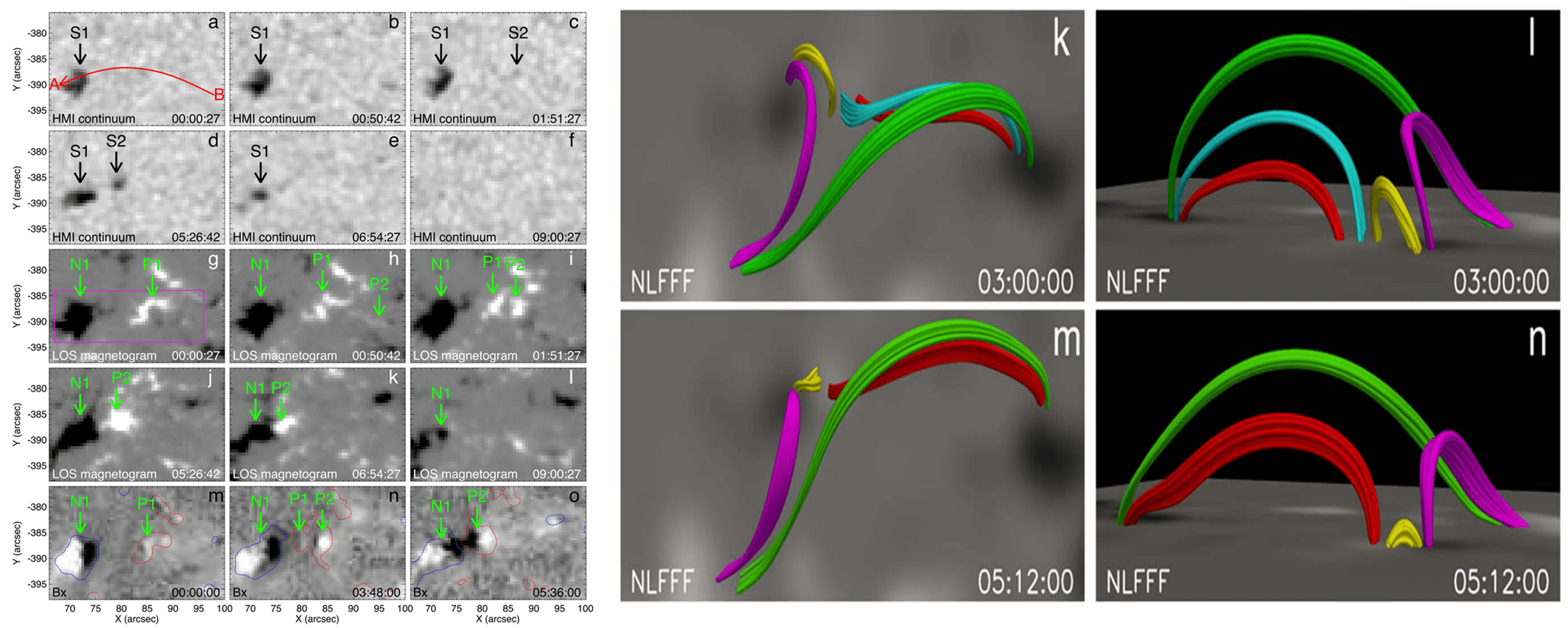}
\caption{Evolution of decaying of pores driven by small-scale magnetic reconnection, as shown by \cite{2021ApJ...919L..29X} in data from SDO on 2020 October 26. Left panels show sequence of continuum images, LOS magnetograms, and Bx, as labeled, displaying the evolution of positive/negative white/black magnetic kernels (N1, P1, and P2). Right panels (k,l,m,n) are snapshots of a numerical simulation showing 3D NLFFF configurations during the magnetic reconnection events, with loops before reconnection takes place (pink), and shorter/longer (yellow/green) loops formed during the reconnection process.}
\label{fig:fig11}
\end{figure}

As mentioned earlier, magnetic field is constantly removed from the solar surface and replenished by the creation of new magnetic field visible via the flux emergence on the solar surface.
It is believed that magnetic fields from large scales, like sunspots, are broken up and shredded by turbulent motions also called turbulent diffusion. Important observational constraints are the moat flow region around a sunspot (a region of out-flowing plasma) leading to the phenomenon of so-called moving magnetic features \cite[MMFs; e.g.][]{Ryutova2018}, which transport magnetic field away from the sunspot. Another characteristic sign for a breaking up of a sunspot is the formation of light-bridges \cite[][]{2003ApJ...589L.117B} which are regions of non suppressed convective flows within sunspots. The shredding process happens until the magnetic energy comes down into sufficient small scales to annihilate with opposite polarity elements \citep[e.g.,][]{2005ApJ...626L.125B}. On these small scales various processes are observed, such as splitting and further dispersal leading ultimately to in-situ disappearance in the case no clear counterpart of the disappearing opposite flux can be seen. However, more often it can be evidenced that the opposite magnetic field polarity approaches a flux element, in a process kind of opposite to the emergence, until the two opposite polarities meet and start to cancel out; first weakening the strong magnetic field elements until they completely disappear \cite[][]{2018ApJ...857...48G}. In the meeting process obviously magnetic reconnection happens on the one or other height, reconnecting the field lines and leading to a new magnetic field topology.
Figure~\ref{fig:fig10}, taken from \cite{2019LRSP...16....1B}, shows three different scenarios: cancellation, disappearance, and fragmentation of magnetic flux features. Depending on the height of the reconnection site (see right panel in Fig.\ref{fig:fig10}) two U-shaped loops are formed with the lower one being pushed down into the convection zone and the upper one being lifted in the higher atmosphere. During such reconnection events, jets and fibrils can be formed, shooting plasma further up. Such reconnection processes also play a fundamental role in the disappearance of magnetic pores as witnessed in \cite[][]{2021ApJ...919L..29X}. These authors observe an active region with forming and decaying pores where the decaying of the pores is due to approaching counter polarities which lead to 2 magnetic reconnection events and a strong change in the magnetic field line topology as indicated by non-linear force free magnetic field (NLFFF) extrapolations during the evolution of the pores, as shown in Fig.~\ref{fig:fig11}.  

\section{Outlook}

New facilities and instrument suites that have recently become available, or are currently in development, will hopefully shed new insights on plasma processes in the lower solar atmosphere. The recent first light by DKIST will be followed by the planned 4-m European Solar Telescope \citep{2019arXiv191208650S,10.1117/12.2234145}, the 8-m Chinese Giant Solar Telescope \citep{2011ASInC...2...31D}, and the 2-m National Large Solar Telescope of India \citep{2011ASInC...2...37H}. Each of these facilities will offer varying instrument suites with the possibility of high temporal and spatial resolutions which will give us unprecedented detail and, thus, improve our understanding of dynamics and small-scale magnetic fields in the solar of the lower solar atmosphere. It is hoped that these facilities and instruments will give more insights on MBPs, umbral dots, jets and fibrils.

With the latest instruments the community is about to reach spatial resolution levels comparable to the photon mean free path in the photosphere. Thus, beyond this point in spatial size, it becomes difficult for features to create enough imprint to evidence observable changes. However, for spectro-polarimetric observations still larger facilities will be sought after, as one needs to distribute the available photons on a 4 dimensional space (2 spatial dimensions, temporal and spectral) and divide them by at least 4 for the 4 polarization states to accurately estimate the magnetic field. However, in practice even more photons are needed as the polarization signals are rather weak and often only leading to signals of a few percent of the total intensity even for strong magnetic fields and circular polarization states. For linear polarization (measurement of the horizontal fields) and hG and below fields it becomes an art to detect enough photons in a sufficiently short time period. Thus, even though the 4-meter class telescopes open new pathways and insights, from the spectro-polarimetric viewpoint we have not yet reached the end of the story, and from the perspective of open science we are still missing the connection between the atmospheric layers as well as simple fundamental questions such as how the field is created on small-scales and how it eventually disappears. In addition, one has to consider the need for inputs from privileged locations in space on the high-resolution front. This is of great importance to address fundamentally important questions regarding the polar magnetic fields in the Sun. Those magnetic field regions are not, or only very limited, accessible from the ecliptic plane due to foreshortening effects. Hence the community still does not understand sufficiently well the magnetic fields of this mysterious region and only with the help of space missions, going considerably out of the ecliptic plane to acquire clear images from the solar poles, those enigmatic regions on the Sun could be fully understood. Assisting us in finally understanding the whole solar cycle as well as to verify the magnetic field dynamo as currently theorized. At present, we have a decent understanding of various aspects of the solar atmosphere, however, there are still many open questions that need to be addressed in the future. The interested reader might like to have a look into the Critical Science Plan of DKIST \citep{2021SoPh..296...70R} or the Science Requirement Document of EST \citep{2019arXiv191208650S}, were many of the to-be-addressed problems are outlined in detail.

Furthermore, a very relevant topic, only limited covered in this short review, is the simulation of the full lower solar atmosphere from the upper convection zone all the way up to the photosphere and chromosphere. However, going into more details of these simulations is beyond the scope of this work and we refer the interested reader to recent works such as \cite[][and references therein]{2007ApJ...665.1469A, 2020LRSP...17....3L,SKIRVIN2022}.

\vspace{3cm}
The authors state that there is no conflict of interest.\\
We express our gratitude to the anonymous referees for their comments that significantly helped to improve the presentation of some of the ideas.

\bibliography{Bibliography.bib}

\end{document}